\documentclass[aps,preprint,a4,superscriptaddress,altaffilletter]{revtex4}

\newcommand{\ben}{\begin{eqnarray}}
\newcommand{\een}{\end{eqnarray}}
\newcommand{\be}{\begin{equation}}
\newcommand{\ee}{\end{equation}}
\newcommand{\n}{\label}
\newcommand{\no}{\noindent}

\usepackage[dvipdf]{epsfig}
\usepackage{color}
\usepackage{float}
\usepackage{subeqn}
\usepackage{hyperref}

\begin{document}

\author{M\'onica I. Forte}\email{forte.monica@gmail.com}
\author{Carlos E. Laciana}
\affiliation{Instituto de Astronom\'ia y F\'isica del Espacio,  1428 Buenos Aires, Argentina}

\title{Can the conformal factor of the gravitational field be a physical degree of freedom?}

\begin{abstract}
The role of the conformal factor was analysed in two gauge-invariant perturbative formulations. Using the classical and quantum linearized perturbation approach given by Mukhanov et al \cite{Mukhanov:1990me}, the non-physical behaviour of the conformal perturbation as an isolated degree of freedom is shown. In the quantum gravity context, the restriction to conformal perturbations gives us an incompatibility with the dynamical equations.
\end{abstract}

\vskip 1cm

\bibliographystyle{plain}
\maketitle

\section{Introduction}

To describe the evolution of the universe near the Planck time, the semiclassical approach, as is well known \cite{Birrell:1982}, is not useful because it is necessary to take into account higher than
1-loop contributions to the effective action. There are also conceptual problems when only
the matter field is quantized; for example, it can be shown \cite{Unruh:1984} by gendanken experiments that
Heisenberg's uncertainty principle is violated. An early attempt to quantize the gravitational
field was performed in \cite{Padmanabhan:1982bd}. There, only one degree of freedom of gravity is quantized; the $\it{conformal\  factor}$. The goal of quantum gravity is to give a probabilistic description of the
evolution of the universe. Then it is necessary to find the probability amplitude to reach a 3-geometry $^{(3)}\mathcal{G}_2$, at the hypersurface $\Sigma_2$, from another initial $^{(3)}\mathcal{G}_1$, at the hypersurface $\Sigma_1$.
The amplitude is denoted by the kernel $ K $ which is formally written as the Feynman functional integral

$$K[^{(3)}\mathcal{G}_2,{\Sigma}_2;^{(3)}\mathcal{G}_1,{\Sigma}_1]=\int \exp(i\mathcal{S}[\mathcal{G}])\mathcal{DG}
$$

In general, the path integral must be performed over all the geometries that satisfy the
boundary conditions. Since this is technically impossible, some restrictions are introduced.
In \cite{Padmanabhan:1982bd} only the metrics produced by a conformal transformation of the initial flat metric are
considered. As is stressed in \cite{Narlikar:1986kr} the physical meaning of that approach is to restrict the metrics
to those that preserve the invariance of the causal relation between two spacetime points,
because the global lightcone structure of a spacetime is preserved in a conformal fluctuation.
Also some phenomenological advantages are obtained with that approach, compared to the
standard cosmology. It is an alternative approach to the inflationary model, and it gives an
explanation of the flatness and the initial singularity problems. In \cite{Padmanabhan:1982bd} it is shown that the
quantization of the conformal degree of freedom is enough to eliminate the initial singularity
of the classical solution of Einstein's equations. In \cite{Padmanabhan:1983pp} it is shown that when the universe is
created through quantum conformal fluctuations, from the Minkowski space, a `fine tuning' of the initial conditions is not necessary. In \cite{Padmanabhan:1989ay} the conformal factor is quantized as a perturbation
of the Friedmann-Robertson-Walker (FRW) metric and the Wheeler-De Witt equation is
obtained, after that the semiclassical limit is performed to obtain the adiabatic approximation.
With respect to the physical meaning of the conformal degree of freedom in these papers it is
argued that it has the same physical status as the scale factor (or the lapse function in the ADM
formalism). In a more recent paper \cite{Souradeep:1992sn}, the conformal perturbations play the role of seeds in
the structure formation.
On the other hand, the energy-momentum contribution to Einstein's equations coming
from the conformal factor is analogous to a scalar field with negative energy. This behaviour
leads to a link with thermo-field dynamics (TFD) (see \cite{Takahasi:1974zn}). In that context the quantum
conformal perturbations can be interpreted as the quantum fluctuations of a thermal reservoir
(tilde modes in TFD) (see \cite{Laciana:1995}). The structure of the Hamiltonian allows us to also relate the
conformal perturbations with the theory of canonical quantum dissipative systems \cite{Alfinito:1997gj}.

However, a strong criticism appears in the context of a rigorous gauge-invariant perturbation theory [1], when one considers the conformal perturbations as physical degrees
of freedom. The object with physical meaning (gauge invariant), related to the conformal
perturbation, is a combination of matter-field perturbations and the conformal perturbation.
The conformal fluctuation by itself, as we will show later, has no physical meaning. A similar
conclusion can be obtained in the quantum gravity context, when a homogeneous and isotropic
background is used in a minisuperspace formalism with gauge-invariant perturbations of the
metric \cite{Halliwell:1984eu}. These theories are not totally covariant because one particular reference system
is assumed at the beginning (the one in which the universe is homogeneous and isotropic), but all the perturbations are considered in a gauge-invariant way.

The organization of this paper is as follows. In section 2 a brief review of the necessary
aspects of the linearized cosmological perturbation approach is given. In section 3 this theory
is applied to the conformal perturbations and their non-physical nature is shown. In section 4
a brief review of the gauge-invariant perturbation formalism in quantum gravity is given. In
section 5 the physical impossibility of having isolated conformal perturbations is shown, due
to the incompatibility with the dynamical equations. Finally, in section 6 the main conclusions
are given.

\section{Review of the linearized gauge-invariant perturbation theory}

From the paper of Mukhanov et al \cite{Mukhanov:1990me}, concerning the theory of cosmological perturbations, we can see the gauge-invariant linearized perturbation formulation. In that approach it is assumed that the perturbations of the spacetime are small deviations of a homogeneous and isotropic spacetime. As is stressed in [1] we must take into account that not all the perturbed metrics correspond to perturbed spacetimes. It is necessary to distinguish  between the changes in the metric due to physical (geometrical) perturbations and those produced by an arbitrary choice of coordinates.

We will show in detail the perturbation of the FRW  metric. In our case the background is given by
\be
\n{1}
ds^2=a^2(d\eta^2-\gamma_{ij}dx^idx^j)
\ee

\no or
\ben
\n{2}
g^{(0)}_{\mu\nu}=a^2 \left( \begin{array}{cc}
1 & 0 \\
0 & -\gamma_{ij} 
\end{array} \right)
\een

The perturbed metric is
\ben
\n{3}
g_{\mu\nu}=g^{(0)}_{\mu\nu}+\delta g_{\mu\nu}
\een

\no where the most general scalar perturbation can be written in the form
\ben
\n{4}
\delta g^{(s)}_{\mu\nu}=a^2 \left( \begin{array}{cc}
2\phi & - B_{|i} \\
- B_{|i} & 2(\psi\gamma_{ij} - E_{|ij} )
\end{array} \right)
\een
From equations (\ref{4}) and (\ref{3}) the line element is
\be
\n{5}
ds^2=a^2(\eta)\{(1+2\phi)d\eta^2- 2B_{|i}d\eta dx^i -[\gamma_{ij}(1-2\psi)+2E_{|ij}]dx^idx^j\}
\ee
\no where the functions $\phi$, $\psi$, B and E give us the deviation with respect to the background metric.

Let us now perform a new perturbation, but due only to a change of coordinates. The most general change of coordinates that preserves the scalar nature of the metric fluctuations is
\ben
\n{6}
  \begin{array}{cc}
\eta \rightarrow \bar\eta=\eta + \xi^0(\eta,x)\\
x^i   \rightarrow \bar{x}^i=x^i +\gamma^{ij}\xi_{|j}(\eta,x)
\end{array} 
\een

\no where  $\xi^0 $ and $\xi$ are two independent functions of the coordinates. Then the change in $\delta g_{\alpha\beta}$ is
\be
\n{7}
\delta  g_{\alpha\beta}\rightarrow \delta \bar g_{\alpha\beta}=\delta g_{\alpha\beta} + \Delta g_{\alpha\beta}
\ee
\no where
\be
\n{8}
\Delta g_{\alpha\beta}= \mathcal{L}_{\xi} g_{\alpha\beta}
\ee

\no is the Lie derivative in the direction of the vector $\xi$ .

Then in the perturbed metric the line element is  
\be
\n{9}
ds^2=a^2(\eta)\{(1+2\bar\phi)d\eta^2- 2\bar B_{|i}d\eta dx^i -[\gamma_{ij}(1-2\bar\psi)+2\bar E_{|ij}]dx^idx^j\}
\ee
\no with

\ben
\n{10}
  \begin{array}{ll}
\bar\phi  =  \phi - \xi^{0'} - \frac{a'}{a}\xi^0 \\
\bar\psi   = \psi  + \frac{a'}{a}\xi^0 \\
    \bar B  =  B + \xi^0 -\xi' \\
    \bar E  =  E -\xi
\end{array} 
\een

\no where ( $'$ ) is the derivative with respect to the conformal time. Equations (\ref{10}) are the gauge transformations. The simplest gauge-invariant quantities, constructed from the metric perturbation, are known as Bardeen invariants. These invariants have the functional form

\ben
\n{11}
 \begin{array}{ll}
\Phi  =  \phi + \frac{1}{a}[(B-E')a]' \\
\Psi = \psi - \frac{a'}{a}(B-E').
\end{array} 
\een

Einstein's equations with matter content are also perturbed, giving a set of equations for the background and another set for the gauge-invariant perturbations. We begin with the unperturbed action in a Robertson-Walker metric, which is given by

\be
\n{12}
S =  - \frac{1}{16\pi g}\int R\sqrt{-g}d^4x + \int (\frac{1}{2}\varphi'_{\alpha}\varphi'^{\alpha} - V(\varphi))\sqrt{-g}d^4x
\ee

Here, it is assumed that $V(\varphi)$ depends only on $\varphi$, therefore the coupling with the curvature
must necessarily be zero. Then we can describe with this a minimally coupled massive field or also a $\lambda \varphi^4$ theory.

Due to the fact that a homogeneous and isotropic universe, with small scalar metric
perturbations, is considered, the scalar field must also be approximately homogeneous and can be decomposed into the form 
$$\varphi(x,t)=\varphi_0(t) + \delta\varphi(x,t).$$

The background Einstein equations are

\ben
\n{13}
 \begin{array}{ll}
\mathcal{H}^2  =  l^2(\frac{1}{2}{\varphi}'^2_0 + V(\varphi_0)a^2) \\
2\mathcal{H}'+\mathcal{H}^2= 3l^2({-\frac{1}{2}\varphi}'^2_0 + V(\varphi_0)a^2).
\end{array} 
\een
\no where $\mathcal{H}\equiv a'/a$ and $l^2\equiv 8\pi G/3$.

In the theory of cosmological perturbations one considers up to second-order perturbation in the action
\be
\n{14}
\delta_2S=\delta_2S_g+\delta_2S_m
\ee

In the paper by Mukhanov et al \cite{Mukhanov:1990me} (we will refer to this paper as the MFB approach) the
gauge-invariant function is also introduced
\be
\n{15}
v=a\left(\delta\varphi+ \frac{\varphi'_0}{\mathcal{H}}\psi\right)
\ee

Rewriting equation (15) with explicit gauge invariance we have

\be
\n{16}
v=a\left(\delta\varphi^{(gi)} + \frac{\varphi'_0}{\mathcal{H}}\Psi\right)
\ee

\no where
\be
\n{17}
\delta\varphi^{gi}=\delta\varphi+ \varphi'_0(B - E').
\ee

In terms of $v$, dropping the surface terms, the perturbation of the action can be written as
\be
\n{18}
\delta_2S=\frac{1}{2}\int \left\{v'^2 - v'_i v'_i +\frac{z''}{z}v^2    \right\}d^4x
\ee

\no with $$z \equiv a \frac{\varphi'_0}{\mathcal{H}}$$

\no where the physical object, analogous to the electric or magnetic fields in electromagnetism, is $v$. This object will play the role of the field in a quantum field theory. Although equation (\ref{18}) has a gauge-invariant form, it is not a covariant expression,  as $ z''/z$ is not covariant.

\section{The conformal perturbation and the MFB approach}

A general scalar perturbation of the metric can be written in the form given by equation (\ref{4}), and the line element for the background and scalar metric perturbations is represented by
equation (\ref{5}). In particular, we now assume a conformal transformation given by 
\be
\n{19}
^{(0)}g_{\mu\nu}\rightarrow g_{\mu\nu}=\ ^{(0)}g_{\mu\nu}\exp u.
\ee

If we consider u as a perturbation, to first order we can write
 \be
\n{20}
g_{\mu\nu} \simeq\ ^{(0)}g_{\mu\nu} + u\ ^{(0)}g_{\mu\nu}. 
\ee

For this case the line element is
\be
\n{21}
ds^2=a^2(\eta)\{(1+u) d\eta^2-(1+u)\gamma_{ij}dx^idx^j\}.
\ee

From the comparison of the latter equation with equation (\ref{5}), we obtain the following conditions:
$$B_{|i}=0 \ \ \ \ \ \ \ E_{|i}= \gamma_{ij}M$$
\no with $M$ a scalar function of the coordinates.

Moreover, to complete the identification, $u$ must satisfy
\be
\n{22}
u=2\phi
\ee

\no and also
\be
\n{23}
u=-2\psi +2M
\ee

From equations (\ref{22}) and (\ref{23}), and the condition on $B$, we obtain the following equations, which give us the  conformal perturbation conditions (CPC):

\begin{subequations}
\be\label{24}
B_{|i}=0
\ee
\be\label{second}
\phi=-\psi +M
\ee
\end{subequations}

Equations (\ref{24}) are not gauge invariant, as we can easily prove using a {\it{reductio ad absurdum}} argument. If we suppose that equation (\ref{second}) is gauge invariant, then the following equation must be valid:
$$\bar\phi=-\bar\psi +\bar M$$
However, when we use the transformation given by equations (\ref{10}) it gives
$$\phi=-\psi + M - \xi + \xi^{0'}  $$

\no in contradiction with equation (\ref{second}). Then, the CPC are broken by a general change of coordinates. The existence or not of conformal fluctuations depends on some particular choice of the coordinate system, therefore the changes in the metrics are not physical. Another manifestation of the loss of gauge invariance can be shown by replacing equation (\ref{23}) in equation (\ref{16}), then we can write
\be
\n{25}
v = a \left\{\delta\varphi^{(gi)} - \frac{\varphi'_0 u}{2\mathcal{H}} + \frac{\varphi'_0}{\mathcal{H}}[M - \mathcal{H}(B - E')]\right\}.
\ee

In one particular gauge (longitudinal) the last term disappears and the function $v$ shows a breaking of the gauge invariance, due to the dependence on $u$ of the second term on the right-hand side of equation (\ref{25}).

As we can see from equation (\ref{25}) it is not possible to obtain gauge-invariant conformal perturbations, while the situation is different for the matter–field perturbation, because when the background  $\varphi_0$  is not time dependent, as we can see from equation (\ref{17}), the field perturbation is gauge invariant. It is also important to note that when the quantization is performed usually \cite{Birrell:1982} the background of the field is zero. In that case,  as we see from equation (\ref{25}), the conformal
perturbation disappears. Then we can say that this perturbation is pure gauge. This result will be obtained again in the context of quantum gravity. We can also see from equation (\ref{25}) that
it is not possible to isolate the conformal perturbation as a gauge-invariant degree of freedom, therefore that object does not have a physical meaning.

\section{Review of the gauge-invariant perturbation theory in quantum gravity}

In the formulation given by Halliwell and Hawking \cite{Halliwell:1984eu} the quantum state of the universe is
described by a wavefunction defined by a path integral on a finite-dimensional space called minisuperspace. That is considered a compact 3-surface $S$ introducing a coordinate $t$ so that
$S$ is the surface $t = 0$. For that foliation of the spacetime, the metric takes the form
\be
\n{26}
ds^2=(N^2 - N_iN^i)d\eta^2- 2N_i dx^i dt - h_{ij}dx^idx^j
\ee

\no with $N$ the lapse function and $N_i$ the shift vector. The 3-geometries obtained as a perturbation of the FRW metric are considered in the path integral. Then, if  equation (\ref{26}) describes the perturbed metric, the function $h_{ij}$ has the form
\be
\n{27}
 h_{ij}= a^2(\Omega_{ij}+\epsilon_{ij})
\ee
and
$$a^2\Omega_{ij}dx^idx^j=a^2[d\chi^2 + \sin^2\chi (\sin^2\theta \ \ d\phi^2 +d\theta^2)]$$

\no in harmonic coordinates.

 $\epsilon_{ij}$ is the perturbation, which in general form is written as
  
\be
\n{28}
\epsilon_{ij}= \sum_{nlm}\left[\frac{1}{3}\sqrt{6}a_{nlm}\Omega_{ij}Q^n_{lm}+\sqrt{6}b_{nlm}(P_{ij})^n_{lm}
+ \sqrt{2}c^0_{nlm}(S^0_{ij})^n_{lm} \right.
\ee
$$\left. \ \ \ \ \ \ \ \ +\sqrt{2}c^e_{nlm}(S^e_{ij})^n_{lm}+2d^0_{nlm}(G^0_{ij})^n_{lm}+2d^e_{nlm}(G^e_{ij})^n_{lm}\right]$$

\no where the coefficients $a_{nlm}$, $b_{nlm}$, $c^0_{nlm}$, $d^0_{nlm}$, are only time-dependent functions. The $Q(x^i)$
are the standard scalar harmonics on the 3-sphere. $P^n_{ij}$, $S^n_{ij}$, $G^n_{ij}$ are harmonic tensors; scalar-, vector- and tensorlike.

In terms of these harmonics we can write the lapse and shift functions as
$$ N = N_0\left[1+\frac{1}{\sqrt{6}}\sum_{nlm}g_{nlm}Q^n_{lm}\right]$$

$$ N_i = \exp \alpha \sum_{nlm}\left[\frac{1}{\sqrt{6}}k_{nlm}(P_i)^n_{lm}+\sqrt{2}j_{nlm}(S_i)^n_{lm}\right]$$

\no where $a(t)=\exp \alpha$ . In the following explanation we will use the reduced notation without indices. In this formulation the coefficients of the expansion of equation (\ref{28}) and $\alpha$ are independent variables, with conjugate momenta $\Pi_\alpha = \partial \mathcal L / \partial \dot\alpha$, $\Pi_{d_n} = \partial \mathcal L / \partial \dot{d_n}$, etc.

Following \cite{Halliwell:1984eu} strictly, we can now write the action in the form
$$S = \int dt (\Pi_q \dot q - H)$$
\no obtaining the Hamiltonian
\be
\n{29}
H = N_0\left[H_{|0}+ \sum_n H^n_{|2}+ \sum_n g_nH^n_{|1}\right]+\sum_n(k_n\  ^sH^n_{-1}+j_n\  ^vH^n_{-1}).
\ee

We can see the explicit expression of each term in \cite{Wada:1986uy}. The term $H_{|0}$  is the Hamiltonian of the unperturbed model with $N=1$. The Hamiltonian $H_{|2}$  is second order in the coefficients  of the spherical harmonics. $H_{|1}$  is first order, $^sH^n_{-1}$  and $^vH^n_{-1}$ are the scalar and vector shift parts of the Hamiltonian.

From the Hamiltonian of equation (\ref{29}) the Schr\"odinger equation is obtained $$\hat H \Psi = 0$$

giving finally
\be
\n{30}
\left[H_{|0}+ \sum_n H^n_{|2}+ \sum_n g_nH^n_{|1}\right]\Psi = 0
\ee
\be
\n{31}
\left[\sum_n(k_n\  ^sH^n_{-1}+j_n\  ^vH^n_{-1})\right]\Psi = 0
\ee

Equation (\ref{30}) is the well known Wheeler-De Witt equation and (\ref{31}) is the supermomentum equation.

\section{Conformal perturbations in the gauge-invariant perturbation formalism of quantum gravity}

We will now use \cite{Hawking:1993tu}. In this paper the relations between the coefficients of the spherical harmonic used in \cite{Halliwell:1984eu} and the functions introduced in [1] are shown. Those relations are

$$ \phi = -\sum_n \frac{g_nQ^n}{\sqrt{6}}$$
$$\psi =\sum_n\frac{(a_n+b_n)Q^n}{\sqrt{6}}$$
$$B = -\sum_n\frac{k_nQ^n}{(n^2-1)\sqrt{6}}$$
$$E = -\sum_n\frac{3b_n Q^n}{(n^2-1)\sqrt{6}}$$

Then $b_n = k_n = 0$  is equivalent to $B = E = 0$ and the line element, replacing in equation (\ref{9}), has the form
\be
\n{32}
ds^2= a^2(\eta)\left\{ \left(1-\sum_n \frac{2g_nQ^n}{\sqrt{6}}\right)d\eta^2   - \left(1-\sum_n \frac{2a_n Q^n}{\sqrt{6}}\right)\gamma_{ij}dx^idx^j \right\}.
\ee

From equation (\ref{32}), for the conformal perturbations it is
\be
\n{33}
g_n =a_n.
\ee

We will now see that the relation (\ref{33}) is incompatible with the dynamics of the fluctuations which is deduced from the Wheeler-De Witt equations.

The equations that give us the dynamics of the classical variables (background) and the fluctuations, can be obtained as an extremum of the action:

\be
\n{34}
I = I_0(\alpha,\phi, N_0)+\sum_n I_n 
\ee
\no where $I_0$ corresponds to the unperturbed system, $\phi$ the matter field,
\be
\n{35}
I_0= -\frac{1}{2}\int dt N_0 e^{3\alpha}\left[\frac{\dot\alpha^2}{N_0^2}-e^{-2\alpha}-\frac{\dot\phi^2}{N_0^2}+m^2 \phi^2\right]
\ee
 
 \no and $I_n$ to the perturbations

$$ I_n=  \int dt (L^n_g +   L^n_m) $$

\no the expressions of $L^n_g$ and $L^n_m$ may be found in \cite{Halliwell:1984eu} (equations (B4) and (B5)).

The variation of $I$ with respect to $b_n$ gives us the equation (see \cite{Hawking:1993tu}, equation (A8)):
 
\be
\n{36}
N_0\frac{d}{dt}\left[ \frac{a^3}{N_0}\frac{db_n}{dt}\right]- \frac{1}{3}(n^2-1)N_0^2a(a_n+b_n)=\frac{1}{3}N_0^2 (n^2-1)ag_n+\frac{1}{3}N_0\frac{d}{dt}\left[\frac{a^2k_n}{N_0}\right].
\ee

Imposing at this level again the condition $b_n = k_n = 0$  for any time (therefore the  derivatives are also null), from equation (\ref{36}) we find

\be
\n{37}
a_n = - g_n.
\ee

This equation is clearly incompatible with equation (\ref{33}) (except for the trivial case when the fluctuations are null). Then the pure conformal perturbations are not physical in this context.

It is interesting to comment on the relation with the work of Wada \cite{Wada:1986uy}. There it is shown that in pure gravity the tensor perturbations (the gravitons) are the only physical degrees of freedom. The Wheeler-De Witt equations can be written as 

\ben
\n{38}
 \begin{array}{ll}
\left[ -\frac{\partial}{\partial a_n}+\frac{1}{3}\{ (n^2-1)a_n+(n^2-4)b_n\}\frac{\partial}{\partial\alpha}\right] \Psi= 0 \\
\\

\left[ -\frac{\partial}{\partial a_n} + \frac{\partial}{\partial b_n} +\left\{a_n + 4 \frac{(n^2-4)}{n^2-1}b_n  \right\} \frac{\partial}{\partial\alpha}\right]\Psi = 0 \\
\\

\left[ -\frac{\partial}{\partial c_n}+4(n^2-4)c_n\frac{\partial}{\partial\alpha}\right]\Psi= 0.
\end{array} 
\een
\\
In \cite{Wada:1986uy} the following change of variable is introduced:
$$\Psi(\alpha, a_n, b_n, c_n, d_n) = \Psi(\tilde\alpha, d_n)$$
\no where

$$\tilde\alpha \equiv  \alpha + \frac{1}{6}\sum_n (n^2-4)(a_n + b_n)^2+\frac{1}{2}\sum_n a_n^2 - 2\sum_n\frac{(n^2-4)}{(n^2-1)} b^2_n $$
$$\ \ \ \ \ \  - 2\sum_n(n^2-4)c^2_n - 2\sum_nd_n.$$

With this transformation all the constraint equations are satisfied automatically and the resulting Wheeler-De Witt equation is 
$$ \left\{H_{|0}(\tilde\alpha)+ \sum_n\ ^T {\tilde H}^n_{|2}(\tilde\alpha,d_n)\right\}\Psi (\tilde\alpha,d_n)= 0$$

\no where $H_{|0} $ is the background Hamiltonian and
$$ ^T {\tilde H}^n_{|2}= \frac{1}{2}\{\Pi^2_{d_n}+(n^2-1)e^{4\tilde\alpha}d^2_n\}$$

Then all the scalar perturbations disappear with an appropriate gauge. This is coincident with the fact that the gravitons are particles of spin two. Therefore, this confirms that the conformal perturbation is pure gauge.

On the other hand, as shown in \cite{Wada:1986uz}, it is not possible to eliminate totally the scalar degrees of freedom coming from the perturbation of the metric, when a scalar field, as matter content, is present. This is also coincident with the result obtained in the MFB approach.

\section{Conclusions}

In the context of the linearized gravitational perturbation \cite{Mukhanov:1990me}, when we consider only conformal fluctuations of the metric and the perturbations of a matter scalar field, it turns out to be impossible to separate both perturbations, as physically independent contributions, by means
of a particular gauge, except when the background of the matter field is a constant $(\varphi'_0= 0)$. However, in that case the conformal perturbation disappears. Therefore, the conformal perturbation is `pure gauge'. It is interesting to show the value of the scale factor for the last example. From the background equations we have
$$\mathcal H^2 - \mathcal H' = \frac{3}{2}l^2{\varphi'}_0^2 $$

\no but when $\varphi'_0=0$, the solution of the background is the inflationary universe $a(t) \propto e^{ \alpha t}$, and $v = a\delta \varphi^{(gi)}$.  For some particular value of $\alpha$ this field can have a behaviour analogous to a massless scalar field, because for this example $m_{ef}= a^2(m^2-\alpha^2)$. 

In the context of quantum gravity with the 3-geometries coming from the gauge-invariant perturbations of a FRW-like background metric, we obtain an inconsistency with the dynamical equations, when the restriction of having only conformal perturbations is imposed. Moreover, in \cite{Wada:1986uy} it is shown that the perturbation contribution in pure gravity is tensorlike (gravitons).
Therefore, also in this context the non-physical nature of the conformal perturbation is proved.

A slightly different approach to the one used by MFB, for the treatment of gauge-invariant perturbations was used in \cite{Miedema:1996uf}. The main difference is that these authors perform a gauge-invariant
perturbation on the energy density and the number of created particles, instead of on the energy-momentum tensor and the Einstein tensor, as in the MFB approach.  In \cite{Miedema:1996uf} the dynamical equation for the gauge-invariant perturbation is identical to Poisson's equation of the Newtonian theory, while the MFB approach differs in terms that are multiplied by the expansion velocity of the universe. Both approaches clearly coincide when we come closer to the actual universe, because the expansion is produced slowly  $(\dot a \sim t_{actual}^{-1/3})$. However, none of the approaches used in \cite{Miedema:1996uf} is in conflict with our results, because the perturbations are separated into three independent groups: gravitational waves (tensorial perturbations), rotational perturbations (vectorial perturbations) and perturbations on the energy density (scalar perturbations). Therefore, the existence of scalar perturbations is only related to the presence of the matter field. Any other kind of scalar perturbations is pure gauge.  Therefore, also in this representation there is no place for the conformal perturbations.

In \cite{Padmanabhan:1983pp}, as we mentioned in the introduction, this study is based on the quantization of
gravity without matter, by means of the path-integral technique on a set of metrics. Those
metrics are only those obtained by conformal transformations from a background classical
metric. The conformal factor is treated as a quantum variable. Therefore, non-physical
paths are considered in the integral and therefore the conclusions are not physically relevant.
The physical object to quantize, instead of the conformal mode, is the field $`v'$ given by equation (\ref{16}). That quantization is performed in \cite{Mukhanov:1990me}, where a fluctuation spectrum in agreement
with the observational data is obtained. The same quantization as in \cite{Padmanabhan:1983pp} is used in \cite{Padmanabhan:1982bd}, however,
it would be interesting to repeat the reasoning, in a forthcoming paper, using gauge-invariant
perturbations instead of conformal ones. It is possible that a similar behaviour to that obtained
in \cite{Padmanabhan:1982bd}, near the singularity, can be obtained, due to the fact that the functional form of the
retarded Green function is in our case also related to a Klein-Gordon-like operator.
In \cite{Souradeep:1992sn} the variation of the gravitational action with respect only to the conformal mode
is performed, then not all the Einstein equations are considered, so the dynamics is taken
into account in an incomplete way. Moreover, as the conformal fluctuations are non-physical
degrees of freedom it is meaningless to think about those fluctuations as seed perturbations
of the universe. However, it would be interesting to study whether that role can be played by the
gauge-invariant fluctuations of the metric using the formalism of \cite{Mukhanov:1990me}.
The non-physical behaviour of the conformal degree of freedom is related to the fact that
it is not possible to add in an $ad$  $hoc$ way a new scalar degree of freedom to the metric tensor.
All the physical changes can be produced if the tensorial nature is not affected. Then only
tensorial perturbations are related to pure gravity, as is shown in \cite{Wada:1986uy}. We can think that the
scalar field introduces some characteristic length in the universe such as an `elemental particle radius', while in pure gravity the conformal perturbation is a mere change of scale.

\section*{ACKNOWLEDGMENTS}

CEL thanks Professor Edgard Gunzig for his hospitality and interesting comments during his visit to the {\it{Universite Libre de Bruxelles}}. This work was partially supported by EEC grants numbers PSS 0992 and CI-CT94-0004, and by OLAM, Fondation pour la Recherche Fondamental, Brussels.


\begin{thebibliography}{99}


\bibitem{Mukhanov:1990me}
  V.~F.~Mukhanov, H.~A.~Feldman, R.~H.~Brandenberger,
  Phys.\ Rept.\  {\bf 215 } (1992)  203-333.
  

\bibitem{Birrell:1982}
Birrell N.~D. and Davies P.~C.~W.~1982 "Quantum Fields in Curved Space" (Cambridge: Cambridge University Press)

\bibitem{Unruh:1984}
Unruh W.~G.~1984 Steps toward a quantum theory of gravity "Quantum Theory of Gravity" ed. S M Christensen
p.~235 (essays in honour of the 60th birthday of Bryce S De Witt)



\bibitem{Padmanabhan:1982bd}
  T.~Padmanabhan, J.~V.~Narlikar,
  Nature {\bf 295 } (1982)  677-678.
  


\bibitem{Narlikar:1986kr}
  J.~V.~Narlikar, T.~Padmanabhan,
  Dordrecht, Netherlands: Reidel ( 1986) 468 P. ( Fundamental Theories Of Physics).


\bibitem{Padmanabhan:1983pp}
  T.~Padmanabhan,
  Phys.\ Lett.\  {\bf A96 } (1983)  110-112.


\bibitem{Padmanabhan:1989ay}
  T.~Padmanabhan,
  Int.\ J.\ Mod.\ Phys.\  {\bf A4 } (1989)  4735-4818.


\bibitem{Souradeep:1992sn}
  T.~Souradeep,
  Astrophys.\ J.\  {\bf 402 } (1993)  375-381.
  [gr-qc/9208003].


\bibitem{Takahasi:1974zn}
Y.~Takahasi, H.~Umezawa,
  Umezawa H 1995 "Advanced Field Theory: Micro, Macro and Thermal Physics" (New York: AIP)


\bibitem{Laciana:1995}
Laciana C E 1995 Physica A 216 511–517


\bibitem{Alfinito:1997gj}
  E.~Alfinito, R.~Manka, G.~Vitiello,



\bibitem{Halliwell:1984eu}
  J.~J.~Halliwell, S.~W.~Hawking,
  Phys.\ Rev.\  {\bf D31 } (1985)  1777.
  


\bibitem{Hawking:1993tu}
  S.~W.~Hawking, R.~Laflamme, G.~W.~Lyons,
  Phys.\ Rev.\  {\bf D47 } (1993)  5342-5356.
  [gr-qc/9301017].


\bibitem{Wada:1986uy}
  S.~Wada,
  Nucl.\ Phys.\  {\bf B276 } (1986)  729.
  


\bibitem{Wada:1986uz}
  S.~Wada,
  Phys.\ Rev.\  {\bf D34 } (1986)  2272.
  


\bibitem{Miedema:1996uf}
  P.~G.~Miedema, W.~A.~van Leeuwen,
  Phys.\ Rev.\  {\bf D54 } (1996)  7227-7236.
  



\end{thebibliography}
\end{document}